\documentclass{aa}  

\usepackage{graphicx}
\usepackage{lipsum} 
\usepackage{txfonts}
\usepackage{amssymb}
\usepackage{amsmath}
\usepackage{sidecap}
\usepackage{multirow}
\begin{document}

   \title{Detection of scattered light from the hot dust in HD 172555
   \thanks{Based on data collected at the European Southern Observatory, Chile under program 095.C-192}
   \thanks{The reduced images as FITS files are available in electronic form at the CDS via anonymous ftp to cdsarc.u-strasbg.fr (130.79.128.5) or via http://cdsweb.u-strasbg.fr/cgi-bin/qcat?J/A+A/} }

   \author{N.~Engler\inst{\ref{instch1}} 
   \and H.M.~Schmid\inst{\ref{instch1}} 
   \and S.P.~Quanz\inst{\ref{instch1}} 
   \and H.~Avenhaus\inst{\ref{instd1}}
   \and A.~Bazzon\inst{\ref{instch1}}  
   }
   
\institute{
ETH Zurich, Institute for Particle Physics and Astrophysics, 
Wolfgang-Pauli-Strasse 27, 
CH-8093 Zurich, Switzerland \\ \email{englern@phys.ethz.ch}\label{instch1}
\and
Max Planck Institute for Astronomy, K\"{o}nigstuhl 17, 69117
Heidelberg, Germany\label{instd1}
     }

   \date{Received ...; accepted ...}

\abstract
{Debris disks or belts are important signposts for the presence 
of colliding planetesimals and, therefore, for ongoing 
planet formation and evolution processes in young planetary systems.  
Imaging of debris material at small separations from the 
star is very challenging but provides valuable insights into 
the spatial distribution of so-called hot dust produced by solid bodies 
located in or near the habitable zone. We report the
first detection of scattered light from the hot dust around the nearby ($d=28.33$~pc)
A star HD 172555.} 
{We want to constrain the geometric
structure of the detected debris disk using polarimetric differential imaging (PDI) with a spatial resolution of 25~mas and an inner working angle of about 0.1$''$.}
{We measured the polarized light of HD 172555,
with SPHERE-ZIMPOL, in the very broad band (VBB) or RI filter 
($\lambda_c=735$~nm, $\Delta\lambda=290$~nm) for the projected 
separations between $0.08''$ (2.3 au) and $0.77''$ (22 au). 
We constrained the disk parameters by fitting models for
scattering of an optically thin dust disk taking the limited spatial resolution and coronagraphic attenuation 
of our data into account.}
{The geometric structure of the disk in polarized light 
shows roughly the same orientation and outer extent as 
obtained from thermal emission at 18 $\mu$m. Our image 
indicates the presence of a strongly inclined ($i\approx 103.5^\circ$), 
roughly axisymmetric dust belt with an outer radius in the range between 
0.3$''$ (8.5 au) and 0.4$''$ (11.3 au). An inner disk edge is not detected in the data. 
We derive a lower limit
for the polarized flux contrast ratio for the disk of 
$(F_{\rm pol})_{\rm disk}/F_{\rm \ast}> (6.2 \pm 0.6)\cdot 10^{-5}$ in the VBB filter. 
This ratio is small, only $\sim$9~\%, when compared to 
the fractional infrared flux excess ($\approx 7.2\cdot 10^{-4}$). 
The model simulations show that more polarized light could
be produced by the dust located inside $\approx 2$ au, which
cannot be detected with the instrument configuration used.}  
{Our data confirm previous infrared imaging and provide a higher resolution map of the system, which could be further improved with future observations.}



\authorrunning{Engler et al.}

\titlerunning{HD 172555 hot debris disk}

   \maketitle
%

\section{Introduction}

Young stars are often surrounded by circumstellar dust debris disks 
or rings. These consist of solid bodies, such as planetesimals and 
comets, as well as large amounts of dust and small amounts of gas. 
Small dust grains with a broad size distribution are 
generated in steady collisions of solid bodies and perhaps by
the evaporation of comets. Debris disks are usually recognized
by infrared (IR) excess on top of the spectral 
energy distribution (SED) of the stellar photosphere because of the thermal emission 
of the heated dust grains 
\citep[see, e.g.,][for a review]{Wyatt2008, Matthews2014}.
In scattered light, debris disks 
are usually very faint, i.e., $> 10^3$ times fainter than the host star, and 
therefore they are difficult to image.  
To date, several dozen debris disks have been spatially resolved in various wavelength bands, from visible
to millimeter,   
using various space and ground-based telescopes. These data 
provide important constraints on the debris disk morphologies 
\citep[e.g.,][]{Moerchen2010, Schneider2014, 
Choquet2016, Olofsson2016, Bonnefoy2017}.

In the absence of imaging data, the location of the bulk of 
dust in the system can be inferred from the grain temperature 
derived by SED modeling. Based on modeling results, most  debris disks 
have been found to harbor warm dust, meaning that the 
temperature of small grains lies between 100 and 
approximately 300 K 
\citep[e.g.,][]{Moor2006, Trilling2008, Chen2011, Morales2011}. 
Cold debris reside far away from the host star where their 
temperature does not exceed 100 K. The prominent analogs to warm and cold dust belts are the main 
asteroid belt at 2-3.5 au and the Kuiper belt between 
30 and 48 au in the solar system \citep{Wyatt2008}. 
Aside from this, several stars with hot disks have been found 
\citep{Wyatt2007, Fujiwara2009} based on the 
IR excess with an estimate temperature above 300 K. 
The stars that possess 
hot dust are very interesting targets because the submicron-sized
grains at a small small distance from a star could be a signpost for 
transient collision events. HD 172555 (HIP 92024, HR 7012) 
is one of the most studied objects among these special stellar systems
\citep{Cote1987, Schuetz2005, Chen2006, Wyatt2007a, Lisse2009, Smith2012, Riviere-Marichalar2012, Johnson2012, Kiefer2014, Wilson2016, Grady2018}.

\begin{table*} 
      \caption[]{Log of observations with the atmospheric conditions for each run.}
      \centering
         \label{Settings}
                \renewcommand{\arraystretch}{1.3}
         \begin{tabular}{cccccccccc}
            \hline 
            \hline
            Date&Observation & Field &  & \multicolumn{4}{c}{Observing conditions (on average)} \\\cline{5-8} 
            &identification$^1$ & offset [$^\circ$] &  & Airmass & Seeing ["] & Coherence time [ms] & Wind speed [ms$^{-1}$] \\
            \hline
            \hline
            \noalign{\smallskip}
            2015-06-21&OBS172\_0004-0036 & 0 &  &1.32&  1.39 &1.1  & 7.9\\
            2015-06-26&OBS177\_0010-0042 & 120 & &1.31 & 0.78 &3.5  & 6.7\\
            2015-07-12&OBS193\_0049-0081 & 60 & &1.33 & 1.41 &0.9  & 1.8\\
            2015-09-06&OBS249\_0028-0060 & 0 &  &1.35 & 0.93 &3.8  & 9.0\\
            2015-09-06&OBS249\_0061-0094 & 60 &  &1.47 & 1.11 &3.4  & 8.7\\
            2015-09-07&OBS250\_0002-0034 & 120 & &1.55 & 2.01 &1.0  & 16.1\\
            \hline
            \hline
            \noalign{\smallskip}
            
         \end{tabular}

\begin{flushleft} {\bf Notes.} $^{(1)}$ The  observation  identification  corresponds  to  the  fits-file  header  keyword ``origname''  without  prefix ``SPHERE\_ZIMPOL\_''. The first three digits give the day of the year followed by the four-digit observation number. \\
Instrument setup for the deep imaging mode: Fast polarimetry, VBB filter, derotator mode P2, coronagraph V\_CLC\_MT\_WF, and total exposure time for each run 42.7 min. \\
Instrument setup for the flux calibration mode: Fast polarimetry, VBB filter, density filter ND2, derotator mode P2, and total exposure time for each run 14.4 s.\end{flushleft}
   \end{table*}

HD 172555 is a V = 4.8$^{m}$, A7V star \citep{Hog2000, Gray2006} at a distance of 
$28.33\pm 0.19$ pc \citep{GaiaCollaboration2016} and a member of the $\beta$ Pictoris moving group. 
The deduced age for the group is $23\pm3$ Myr \citep{Mamajek2014}. HD 172555 has 
a K5Ve low mass companion CD-64 1208, separated by more 
than 2000~au or 71.4$''$ \citep{Torres2006}. 
The Infrared Astronomical Satellite (IRAS) identified HD 172555 
as a Vega-like star that has a large IR excess with an effective 
temperature of 290~K \citep{Cote1987}. 
Mid-infrared (mid-IR) spectroscopy from the ground \citep{Schuetz2005} and
5.5-35 $\mu$m spectroscopy obtained with the 
\textit{Spitzer} Infrared Spectrograph (IRS) show very pronounced 
SiO features \citep[][their Fig. 2(d)]{Chen2006} 
indicating large amounts of submicron sized crystalline 
silicate grains with high temperatures (> 300 K). \cite{Lisse2009} 
proposed, based on the spectral analysis of mineral features,
that the IR-excess emission originates from a giant collision of 
large rocky planetesimals similar to the event that could have created 
the Earth-Moon system. 

Until now,  only one resolved image of the HD 172555 debris disk 
existed that was obtained in the Qa band
($\lambda_c=18.30\, \mu$m, $\Delta\lambda=1.52\, \mu$m; hereafter Q band) with TReCS 
(Thermal-Region Camera Spectrograph) at Gemini South telescope and described by 
\citet{Smith2012}. They detected extended emission 
after subtraction of a standard star  point spread function (PSF), which is consistent 
with an inclined disk with an orientation of the major axis of $\theta=120^{\circ}$,
an inclination of $\approx75^{\circ}$, and a disk outer radius of 8 au. 
In Si-5 filter ($\lambda_c=11.66\, \mu$m, $\Delta\lambda=1.13\, \mu$m; hereafter N band) data, the disk was not detected. 
Combining these results with the visibility functions measured with the MID-infrared Interferometric instrument (MIDI), \cite{Smith2012} concluded that the warm dust radiating 
at $\sim$10 $\mu$m is located inside 8 au from the star or inside
the detected 18 $\mu$m emission. 

The proximity of the host star and strong excess in the mid-IR
\citep[$L_{\rm {IR}}/L_{\ast}=7.2\cdot 10^{-4}$;][]{Mittal2015} from hot dust, makes HD 172555 an 
excellent, yet challenging, object for the search of
scattered light. In this paper, we report the first detection of scattered light 
from the hot debris disk around HD 172555 with polarimetric differential imaging (PDI).

We present the PDI data from 2015 taken with  
ZIMPOL (Zurich IMaging POLarimeter; Schmid et al. 2018, submitted) at the Very Large Telescope (VLT) in Chile. The ZIMPOL 
instrument provides high contrast imaging polarimetry, coronagraphy with
a small inner working angle of $\sim$0.1$''$, and high spatial 
resolution of $\sim$25 mas. The next section 
presents the available data together with a brief description of 
the instrument. Section 3 describes the data reduction and Section 4
the obtained results. Section 5 includes analysis of the observed disk morphology and photometry. 
For the interpretation of the data we calculated three-dimensional (3D) disk 
models for spatial distribution of dust to put constraints on the geometric parameters 
of the disk and to estimate the flux cancellation effects in the
Stokes $Q$ and $U$ parameters. In Sect.~\ref{Discussion}, we compare our results
with previous TReCS and MIDI observations \citep{Smith2012}. 
We also investigate the possible presence 
of a very compact dust scattering component that would not
be detectable in our data. We conclude and summarize our findings in Sect. \ref{s_Summary}.


\section{Observations} \label{Observations}

Observations of HD 172555 were taken in service or queue mode
during six runs in 2015 as part of the open time
program 095.C-192 using ZIMPOL, the visual subsystem of the SPHERE 
(Spectro-Polarimetric High-contrast Exoplanet REsearch) instrument. 
The SPHERE instrument was developed for high contrast
observations in the visual and near-IR spectral range and includes 
an extreme adaptive optics (AO) system and three focal
plane instruments for differential imaging \citep[][Schmid et al. 2018, submitted]{Beuzit2008,
Kasper2012,Dohlen2006,Fusco2014}. 

The ZIMPOL polarimetry is based on a modulation-demodulation 
technique and our observations of HD 172555 in the VBB filter were carried out in the fast modulation mode using field stabilization, which is also known as the P2 derotator mode.
The ZIMPOL modulation technique measures the opposite polarization
states of incoming light quasi-simultaneously and with the
same pixels such that differential aberrations between $I_\perp$ and
$I_\parallel$ are minimized. Fast modulation means a polarimetric 
cycle frequency of 967.5~Hz, which is obtained with a ferro-electric
liquid crystal (FLC) and a polarization beam splitter in front of 
the demodulating CCD detector. The high detector gain of 10.5 e$^-$/ADU 
allows broadband observations with high photon rates without 
pixel saturation and ensures high sensitivity at small 
angular separations at the expense of a relative high readout noise. 
The  ZIMPOL VBB or RI filter used ($\lambda_c=735$~nm, $\Delta\lambda=290$~nm)
covers the R band and I band. Polarimetric data were taken
in coronagraphic mode using the Lyot coronagraph V\_CLC\_MT\_WF with 
a diameter of 155 mas. The mask is semi-transparent such that the
central star is visible as faint PSF in the middle of attenuation
region if the stellar PSF is of good quality (good AO corrections) and well centered
on the mask. Short flux calibrations (the star was offset from the coronagraphic
mask) with detector integration time (DIT) of 1.2 s and total
exposure time of 14.4~s were also taken
in each run using the neutral density filter ND2 with a transmission of about 0.87\% for the VBB filter to avoid heavy saturation of the PSF peak. 

The ZIMPOL instrument is equipped with two detectors/cameras, cam1 and cam2, 
which both have essentially the same field of view of 
$3.6''\times 3.6''$. The format of the processed frame is $1024\times 1024$ pixels, where 
one pixel corresponds to $3.6\times 3.6$ mas on sky. The 
spatial resolution of our data 
provided by the AO-system is about $25$~mas. 

The HD 172555 data were taken at three different sky orientations 
on the CCD detectors with position angle (PA)\ offsets of 
$0^{\circ},\ 60^{\circ}$ and $120^{\circ}$ with respect to sky north. 
For each field orientation, eight polarimetric QU cycles were 
recorded consisting each of four consecutive exposures of the 
Stokes linear polarization parameters $+Q, -Q, +U$ and $-U$
using half-wave plate (HWP) offset angles of 0$^{\circ}$, 
45$^\circ$, 22.5$^\circ$, and 67.5$^\circ$, respectively.  
For each exposure, eight frames with  DIT 
of 10~s were taken and this yields a total integration of 
10~s $\times$ 8 frames $\times$ 4 exposures $\times$ 8 cycles 
or $\sim$42.7 minutes. Because of a timing error in the control law \citep[see][]{Maire2016} 
of the derotator during the first three
runs performed in June 2015, the program was re-executed in 
August and September 2015. However, it turned out that this timing
error was so small, that it does not  significantly affect the
imaging of an extended source at small separation. Therefore
none of the six disk observations for HD 172555 are degraded by this effect.
All our deep polarimetric observations of HD 172555 are listed
in Table\,\ref{Settings} together with the observing conditions,
which have an important impact on the data quality. 

\section{Data reduction}\label{data}

\begin{figure*} 
   \centering
      \includegraphics[width=17.5cm]{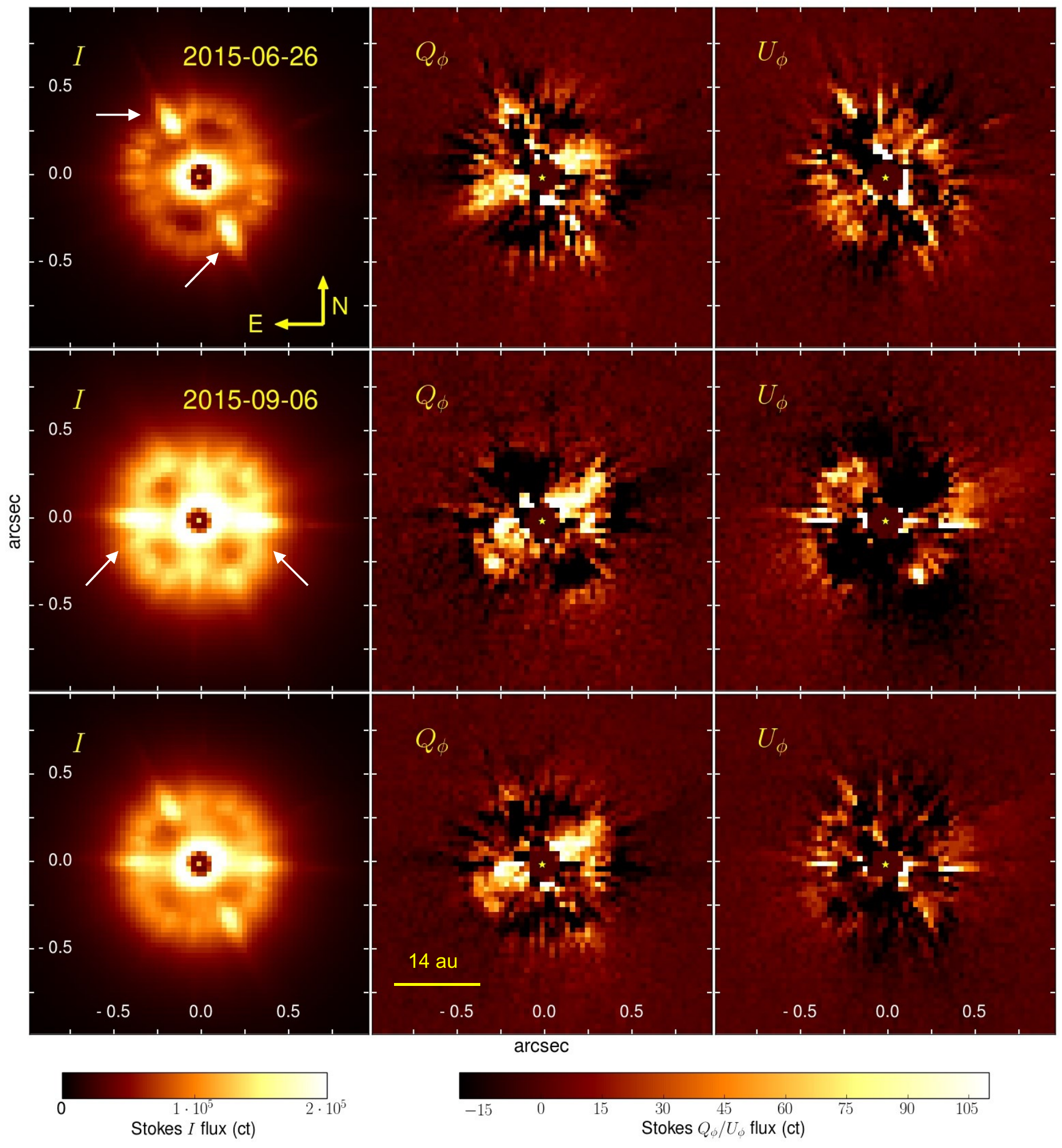}    
                 \caption{Polarimetric differential imaging data of HD 172555 with the VBB filter (590-880 nm). The $8\times 8$ binned images show polarized flux computed as Stokes $I$ (left column), $Q_\varphi$ (middle column), and $U_\varphi$ (right column) parameters. Top row: Observation OBS177\_0010-0042 (see Table \ref{Settings} for the observational conditions) data averaged over 8 QU cycles. Middle row: Observation OBS249\_0028-0060 data averaged over 6 QU cycles. Bottom row: Average of the data presented in the top and middle rows. The position of the star is denoted by a yellow asterisk. White arrows point out quasi-static speckles from the AO system. In all images north is up and east is to the left. The color bars show the counts per binned pixel. \label{QphiUphi}    }
 \end{figure*}

The data were reduced with the SPHERE-ZIMPOL data reduction 
pipeline developed at ETH Zurich. We applied very similar procedures as in \citet{Engler2017}
using flux normalization before polarimetric combination of the data.

For high contrast polarimetry at small separations from the star
an accurate centering and the correction of the ZIMPOL 
polarimetric beam-shift are important (Schmid et al. 2018, submitted). 
The stellar position behind the mask should be determined with 
a precision higher than 0.3 pixels, otherwise the final image 
shows strong offset residuals disturbing the detection of 
a faint polarimetric signal. For the fine centering we fitted a 2D Gaussian function using a central 
subimage with a radius of eight pixels containing the 
stellar light spot transmitted through a semitransparent
coronagraphic focal plane mask.   

The output of the data reduction pipeline consists of images of 
the intensity or Stokes $I$ and the Stokes $Q$ and $U$ fluxes,
where $Q=I_0-I_{90}$ is positive for a linear polarization 
in N-S-direction and $U=I_{45}-I_{135}$ is positive for 
a polarization in NE-SW direction. For the data analysis, 
we transformed the $Q$ and $U$ images into a tangential or
azimuthal polarization images $Q_\varphi$ and $U_\varphi$, locally
defined with respect to the central light source. This transformation
is useful for an optically thin circumstellar scattering 
region because the polarization signal is mainly in
the $Q_\varphi$ component, while $U_\varphi$ should be zero 
\citep[e.g.,][see also Appendix \ref{s_modeling_app}]{Schmid2006}.

During the data reduction, it turned out that an accurate 
centering of our HD 172555 images is difficult and not possible for a
substantial portion of our data because several runs 
suffered from mediocre atmospheric conditions with a seeing $>1''$
and therefore strongly reduced AO performance. Because of this
problem only data from the best night 177 (2015-06-26; see Table~\ref{Settings}) with an average seeing of $0.78''$ 
and most data from the second best night 249 (2015-09-06) with seeing of $0.93''$ 
could be used for our high contrast polarimetry. Figure~\ref{IQU} shows the Stokes $I$, 
$Q,$ and $U$ images of these data sets in comparison with the reduced data from the other observing runs
that were affected by poor observing conditions.

\begin{figure*}   

    \includegraphics[width=17cm]{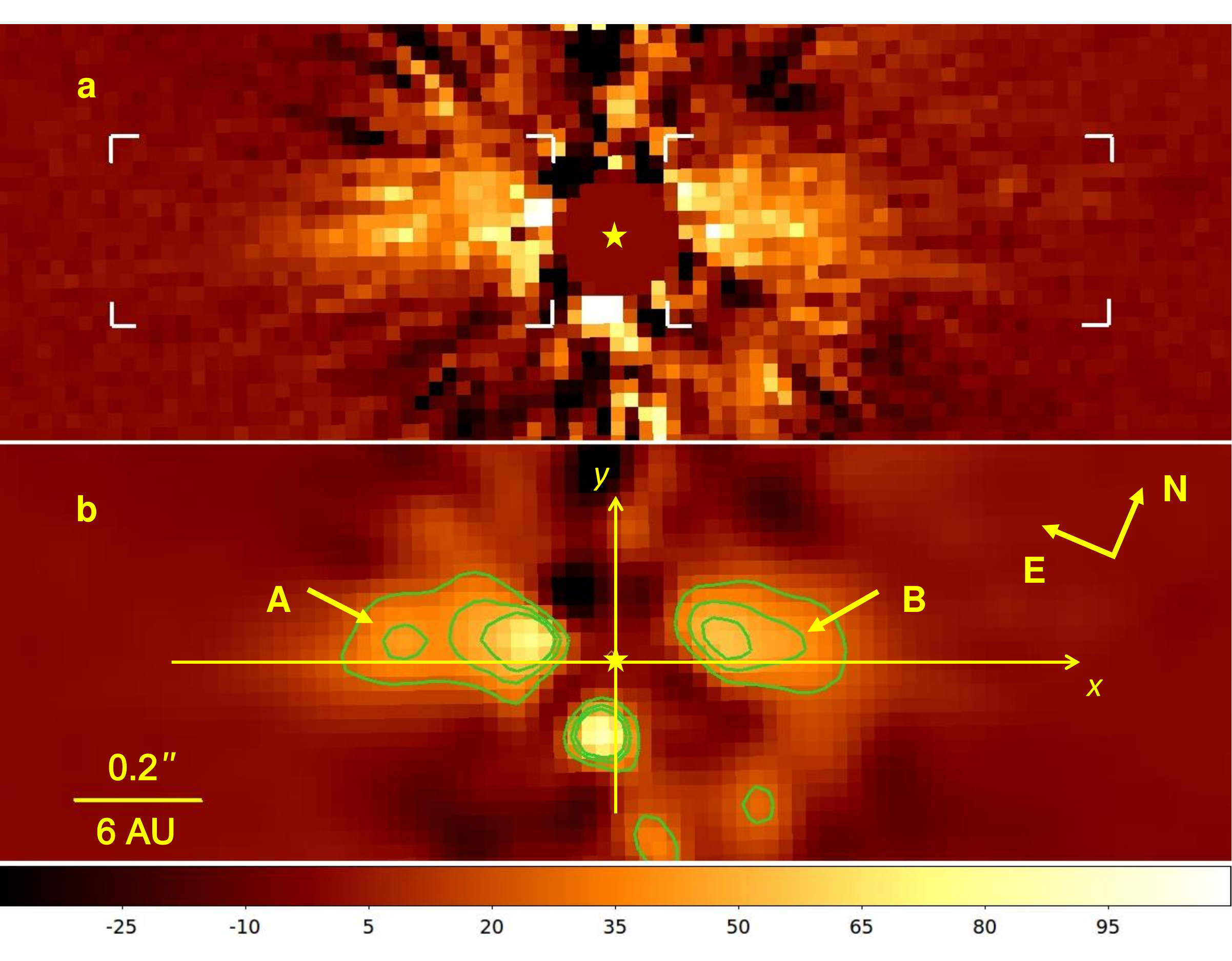} 
   \centering
   \caption{$Q_\varphi$ image (a) and isophotal contours of polarized light intensity overlying the $Q_\varphi$ image (b). The original data (OBS177\_0010-0042) were $6\times 6$ binned to highlight the disk shape and to reduce the photon noise level. The position of the star is indicated by a yellow asterisk. White brackets in the top panel show the boundaries of two rectangular areas where the counts were summed up to retrieve the total polarized flux of the disk (see Sect. \ref{Contrast}). Image (b) is smoothed via a Gaussian kernel with $\sigma_{\rm kernel} =$ 3 px. The isophotal contours show levels of the surface brightness with approx. 25, 40, and 50 counts per binned pixel (from the lowest to the highest contour). Arrows "A" and "B" mark the location of the increased surface brightness, which could indicate the radius of the planetesimal belt. The $x$-axis coincides with the disk axis at estimated PA $\theta_{\rm disk}=112^{\circ}$ (see Sect. \ref{Data Analysis}). The color bar shows the surface brightness in counts per binned pixel. \label{Qphi}}
\end{figure*}

To have a higher S/N per resolution element, images were $8 \times 8$ (Fig. \ref{QphiUphi}) binned for the data analysis and modeling. In Fig.~\ref{Qphi} we used a smaller binning ($6 \times 6$) to preserve the spatial information and to have a more detailed picture of the disk.

\section{Features in the reduced data}\label{s_Results}
The top row of Figure \ref{QphiUphi} shows the main results of
our data reduction, i.e., the mean intensity or Stokes $I$, 
Stokes $Q_\varphi$, and $U_\varphi$ images using all eight cycles
of the best night. The second row shows the same for six out of eight
cycles from the second best night. The bottom row is the mean
of these data.

The quality of the data can be easily recognized in the 
coronagraphic intensity images. These images show, for the best night
(top), a much weaker halo and a weaker speckle ring at a 
separation of about $r\approx 0.35''$ than for the second best night (middle).
For all other nights the disturbing light halo and speckles 
from the central star are even stronger (see Fig.~\ref{IQU}).  
Also clearly visible in Fig.~\ref{QphiUphi} is the rotated speckle pattern in the
top $I$ image, which was taken with a field orientation of $120^\circ$,
with respect to the middle $I$ image taken without field orientation
offset or $0^\circ$. In the latter image two strong quasi-static speckles
from the AO system (indicated with white arrows) are located left and right from the central source
on the speckle ring. The same speckles are rotated in the top
image by 120$^\circ$ because of the applied field rotation. 

The $Q_\varphi$ image from the best night in the top row shows
a positive signal or tangential polarization for regions
just ESE and WNW from the coronagraphic mask and
some noisy features at the location of the two strong speckles.
The corresponding $U_\varphi$ image is also noisy but shows no
predominant extended feature as seen in $Q_\varphi$. The
same ESE--WNW extension is also present in the $Q_\varphi$ image
of the second best night, but now the disturbing strong speckles
are adding noise in the E-W direction, particularly well
visible in the $U_\varphi$ image. Besides this, both $Q_\varphi$
and $U_\varphi$ for the second best night show a quadrant pattern
that is typical for a not perfectly corrected component from the
telescope polarization. The following analysis of the disk morphology 
and modeling are based on the $Q_\varphi$ image 
in the top row obtained under the best observing conditions
because these data are the least affected by noise. 

The fact that we see the same kind of extended
feature in the ESE-WNW orientation 
in both data sets (from the best and second best nights), while other features can clearly be associated with instrumental effects, strongly supports an
interpretation that this is a real detection of the debris dust around
HD 172555. This is further supported by the
fact that the orientation and extent of the detected feature coincides
at least roughly with the disk detection reported previously
by \citet{Smith2012} based on imaging in the mid-IR. 

In our image there is a prominent bright spot located below the
coronagraphic mask. Most likely it is a spurious feature originating from image jitter and the temporal leakage of light of the central star from behind the coronagraphic mask. It is seen more particularly in 177 data set, which argues against this being a real feature.

A final remark on the data quality. Our data look noisy and
are affected by not well corrected instrumental effects. However, 
we note that the signals in the $Q_\varphi$ and $U_\varphi$ images 
are at a very low level of 0.001 of the signal in the intensity frame 
or the surface brightness in the $Q_\varphi$ image is on the order 
$10^{-5}$ of the expected peak flux of the central star
for a separation of $<0.3$~arcsec. The detection of this faint target requires very high contrast
polarimetry at small angular separation. Not all instrumental
effects are understood and can be calibrated at this level of 
sensitivity yet. 

\begin{figure*}
   \centering
  \includegraphics[width=16cm]{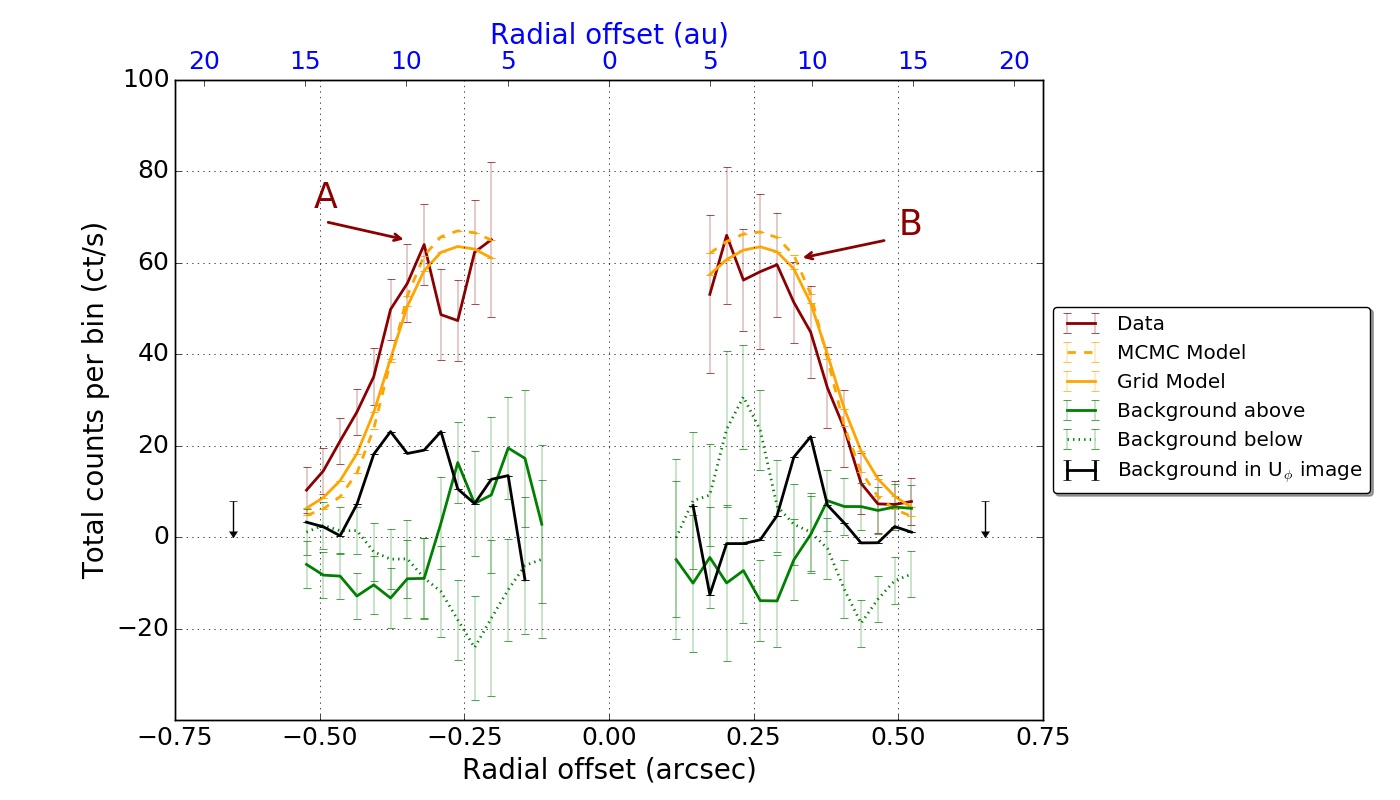} 
   \caption{Radial polarized flux profile measured in the $Q_\varphi$ image per $0.029''(\Delta x) \times 0.290''(\Delta y)$ bin (red solid line) compared with profiles of the best-fitting models and background. The background above (green solid line) and below (light green dotted line) the disk midplane and the background in the $U_\varphi$ image (black solid line) are calculated as explained in Sect.~\ref{Data Analysis}. The upper limit is set for the polarized flux per bin for the disk regions beyond $r= 0.53''$. The arrows "A" and "B" point to the surface brightness peaks. }
    \label{f_SB_profile}
\end{figure*}

\section{Analysis} \label{Data Analysis}
\subsection{Disk position angle and morphology} \label{s_PA}
To determine the PA of the disk $\theta_{\rm disk}$, we varied the position of the disk major axis 
in the range from $\theta_{\rm disk}=105^\circ$ to $\theta_{\rm disk}=125^\circ$ with a step of 0.5$^\circ$. Each time the $Q_\varphi$ image was rotated by $\theta_{\rm disk}-90^{\circ}$ clockwise to place the disk major axis horizontally and then we determined the difference between the left side and the mirrored right side of the image.
The minimum residual flux corresponding to the best match between disk sides is achieved at $\theta_{\rm disk}= 113.5^{\circ}$ for the data set from 2015-06-26 (night 177) and at PA $= 116.5^{\circ}$ for the data set from 2015-09-06 (night 249).

We adopted the PA equal $\theta_{\rm disk}=114^{\circ} \pm 3^{\circ}$. 
This PA should be corrected for the ZIMPOL True North offset of $-2^{\circ}$, which means the preprocessed image
must be rotated by $2^{\circ}$ clockwise. The final PA of the disk is, therefore,
$\theta_{\rm disk}=112^{\circ} \pm 3^{\circ}$ north over east (NoE), 
where the indicated uncertainty includes associated systematic errors. 
For the following data analysis, the $Q_\varphi$ image 
was rotated by $24^{\circ}$ clockwise, including the correction for the True North offset, to place the disk major axis horizontally. 
We introduced an $x-y$ disk coordinate system with the star at  
the origin and the $+x$ coordinate along the derived disk major axis 
in WNW direction and the $y$ coordinate perpendicular to this 
with the positive axis toward NNE ($\theta =22^\circ$) as shown in Fig. \ref{Qphi}(b). 

The observed $Q_\varphi$ flux image shows areas of enhanced surface 
brightness at the separations $|x|\simeq 0.1-0.3''$ but located
about $0.04''$ above the $x$-axis and this up-down asymmetry is reflected clearly
in the surface brightness contours shown in Fig. \ref{Qphi}(b).

\begin{table*}  
      \caption[]{Disk model parameters. \label{t_results}}
             \centering
         \begin{tabular}{l|cccc|cc}
            \hline
            \hline
            \noalign{\smallskip}
      \multirow{ 2}{*}{Optimized parameter}  &\multicolumn{4}{c|}{Grid model} & \multicolumn{2}{c}{MCMC model} \\  

             & Range & Step & Min $\chi^2_\nu$ & Mean value & Range & Best-fit value \\
            \hline
            \noalign{\smallskip}
Position angle $\theta_{\rm disk}$, deg & (...) & (...) & (...) & (...) & [109, 114]& 112.3$^{+1.5}_{-1.5}$ \\[5pt]

Radius of 'belt' $r_0$, arcsec (au) & [0.21, 0.42] & 0.0175 & 0.38 & 0.36$^{+0.06}_{-0.06}$ $\left(10.3^{+1.7}_{-1.7}\right)$ &[0.25, 0.50]& 0.40$^{+0.06}_{-0.06}$  $\left(11.3^{+1.7}_{-1.7}\right)$ \\[5pt]

Scale height $H_0$, arcsec (au) &[0.00, 0.1] & 0.02 & 0.04 & 0.05$^{+0.04}_{-0.04}$ $\left(1.36^{+1.23}_{-1.23}\right)$ &[0.00, 0.1]& 0.02$^{+0.01}_{-0.01}$   $\left(0.6^{+0.3}_{-0.3}\right)$  \\[5pt]

Inner radial index $\alpha_{\rm in}$ &[0, 6] &  1 & 3 & 3.5$^{+3.3}_{-3.3}$  & [0, 6]& 2.3$^{+0.8}_{-0.6}$   \\[5pt]

Outer radial index $\alpha_{\rm out}$ & [-11, 0] & 1 & -7 & -6.8$^{+5.7}_{-5.7}$ & [-20, 0]& -9.8$^{+2.5}_{-4.1}$  \\[5pt]

Flare index $\beta$  & [0.25, 1.50]  & 0.25 & 0.25 & 0.45$^{+2.75}_{-2.75}$ & [0.0, 1.0] & 0.4$^{+0.4}_{-0.3}$  \\[5pt]

Inclination $i$, deg & [96, 111] & 1 & 105 & 103.5$^{+8.2}_{-8.2}$  &[96, 111]& 103.8$^{+1.7}_{-1.7}$  \\[5pt]

HG parameter $g_{\rm sca}$ & [0.2, 1.0]& 0.1 & 0.6 & $0.64^{+0.38}_{-0.38}$ & [0.2, 1.0] & $0.74^{+0.08}_{-0.10}$  \\[5pt]

Scaling factor $A_p$ & (...) & (...) & 1.1 & $0.9^{+12.36}_{-12.36}$   & (...) & $2.8^{+0.8}_{-0.9}$   \\[5pt]
           \noalign{\smallskip}
           \hline
            \hline
            \noalign{\smallskip}
\end{tabular}
\end{table*}

   \begin{figure}[h]
   \centering
   \includegraphics[width=8cm]{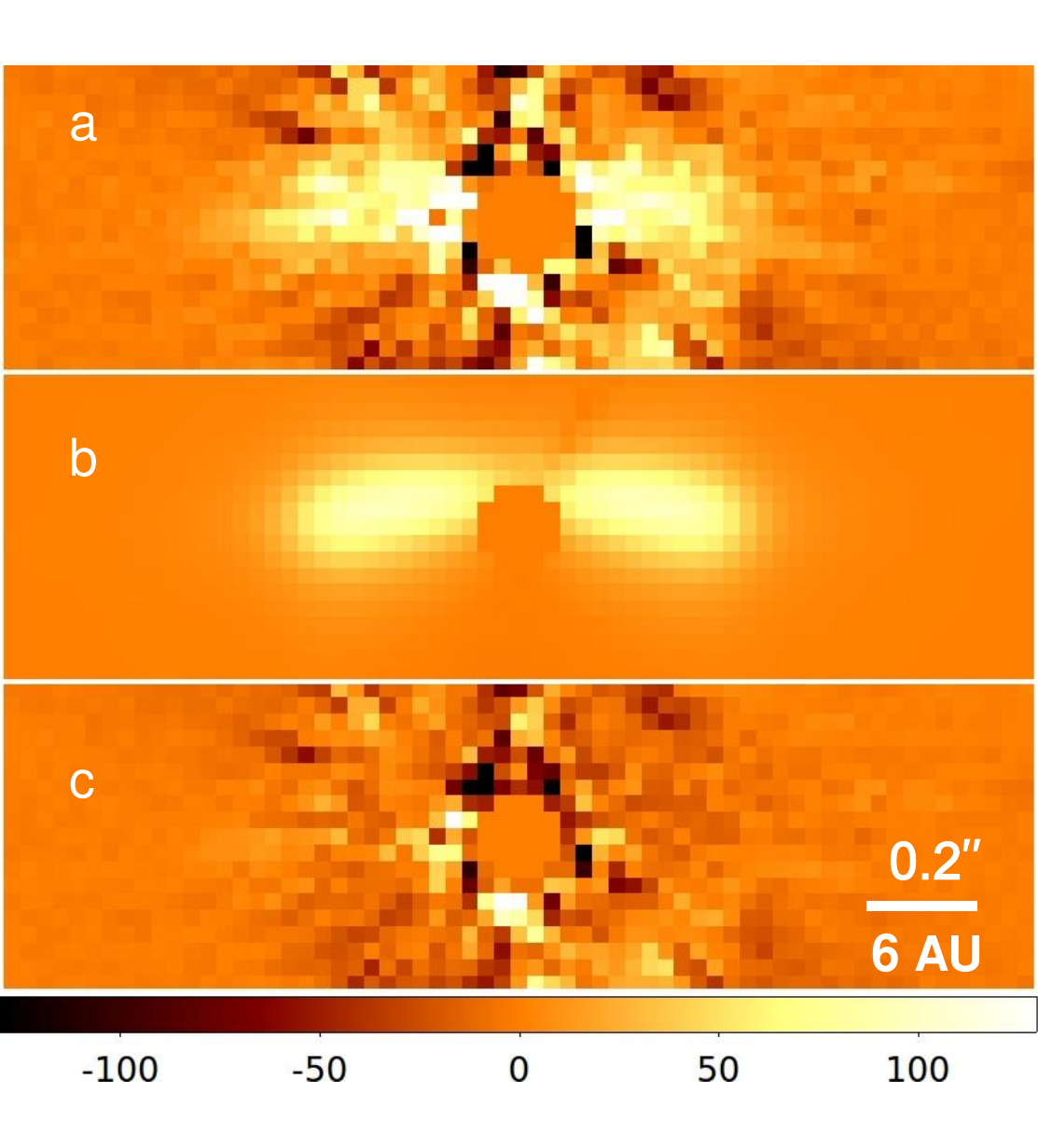} 
   \caption{Comparison of the grid model with the $Q_\varphi$ image. 
(\textbf{a}): $Q_\varphi$ image after 8x8 binning. (\textbf{b}): Image of the best-fitting grid model convolved with the instrumental PSF. (\textbf{c}): Residual image obtained after subtraction of the PSF-convolved grid model image ({\bf b}) from the $Q_\varphi$ image ({\bf a}). Color bar shows flux in counts per binned pixel.  \label{model}}
   \end{figure}  

Figure \ref{f_SB_profile} shows the profile for the polarized flux along the $x$-axis measured
in bins of $\Delta x=29$~mas and integrated in the $y$-direction from
$ y= -0.13''$ to $y=+0.16''$.
Our data show a higher flux closer to the star with a maximum flux inside $r=0.3''$ and 
two small brightness peaks at $|x|\approx 0.3''$ (8.5 au) indicated with arrows "A" and "B" in Figs.~\ref{Qphi}(b) and \ref{f_SB_profile}. 

Figure \ref{f_SB_profile} also includes background profiles determined 
below and above the disk major axis by summing up the counts 
from $ y= -0.42''$ to $y=-0.13''$ (light green dotted line) and from 
$ y= +0.16''$ to $y=+0.45''$ (green solid line). 
The disk flux is at least twice as high as the background, 
except for a small region around $x\approx +0.23''$ where the 
background is particularly high. The origin of the enhanced background below the disk midplane is unclear. The errors on flux per bin were calculated from the variance of the sum of pixel values in bins assuming a sample of normal random variables. The uncertainty of the pixel value is given by the standard deviation of the flux distribution (counts per pixel) computed in concentric annuli in the $Q_\varphi$ image (excluding the area containing the disk flux) to take into account the radial dependence of the data variance. As an additional noise estimator, we measured the background in the $U_\varphi$ image (black solid line in Fig.~\ref{f_SB_profile}) in the same areas where the polarized flux of the disk was measured in the $Q_\varphi$ image.
The polarized scattered light from the disk is detected up to a 
radial distance of $r\sim 0.8''$. Beyond $r=0.3''$, the polarized flux per x-bin is less than 8 ct/s.

\subsection{Modeling the observed disk flux} \label{Modelling}
We modeled the spatial distribution of dust in HD 172555 disk with the 3D 
single scattering code presented in \citet{Engler2017}. 
The disk geometry is described by a radius $r_0$, where the 
radial distribution of the grain number density has a maximum, and
by inner and outer radial power-law falloffs with exponents 
$\alpha_{\rm in}$ and $\alpha_{\rm out}$, respectively. 
For the vertical distribution we adopt the simpler Lorentzian 
profile instead of the exponential profile.
The former profile has already been used for this purpose \citep[e.g.,][]{Krist2005} and has an advantage that it is described with one parameter less. 
The Lorentzian profile is given by
\[f_L(h)=a_L{\left[ 1 + \left(\frac{h}{H(r)}\right)^2\right]}^{-1} ,\]
where $h$ is the height above the disk midplane and $a_L$ is 
the peak number density of grains in the disk midplane. 
The scale height of the disk $H(r)$ is defined as a half width 
at half maximum of the vertical profile at radial distance 
$r$ and scales as in $H(r)=H_0 \, (r/r_0)^\beta$. \\ 

The parameters of the model are the PA of the disk $\theta_{\rm disk}$, the radius of 
the planetesimal belt $r_0$, inner $\alpha_{\rm in}$ and outer $\alpha_{\rm out}$ power law
exponents, scale height $H_0$, flare exponent $\beta$, 
inclination angle $i$, Henyey-Greenstein (HG) scattering asymmetry parameter $g_{\rm sca}$ 
and the scaling factor $A_p$. The scattering phase function for the polarized flux $f_{\rm p} \, (\theta, g_{\rm sca})$ is the product $f_{\rm p} \, (\theta, g_{\rm sca}) = HG (\theta, g_{\rm sca})\cdot p_{\rm Ray}(\theta)$ of the HG function and the Rayleigh scattering function (see Fig.~8 in \citet{Engler2017}).

We assume that the debris disk is optically thin and neglect multiple scattering.
The HG parameter has only positive values ($0\leqslant g_{\rm sca} \leqslant 1 $) 
assuming that the dust particles are predominately forward-scattering
as inferred from imaging polarimetry of other highly inclined 
debris disks \citep[e.g.,][]{Olofsson2016, Engler2017}. This means that the brighter side of the disk is the near side and the inclination higher than 90$^\circ$ defines a disk with its near side toward north as in the case of the HD 172555 disk. 

To simplify the model computation, we adopt the same grain properties everywhere in the disk.
The average scattering cross section per particle is 
a free parameter included in the scaling factor $A_p$.

The $Q_\varphi$ model image is calculated from the combination of the PSF convolved images of the Stokes parameters $Q$ and $U$. The individual steps of the modeling procedure are described and illustrated in Appendix \ref{s_modeling_app}. 

To find parameters of the model that best fits the $Q_\varphi$ image (Fig.~\ref{model}a), we took a twofold approach: first, we fitted the data with models drawn from a parameter grid and second we used the Markov chain Monte Carlo (MCMC) technique. 

\subsubsection{Grid of models} \label{s_Grid}
We define an extensive parameter grid of $\sim$10$^4$ models and calculate a synthetic $Q_\varphi$ image for each model. The explored parameter space and steps of linear sampling are given in Table~\ref{t_results}, Cols. 2 and 3. In this first modeling approach, the disk PA is set to $\theta_{\rm disk}=112^\circ$, which is the PA value found in the previous section (Sect.~\ref{s_PA}).
The flux scaling factor $A_p$ is treated as a free parameter for each model and is obtained by minimizing the $\chi^2$ for the fit of the $Q_\varphi$ model image to the data.
The reduced $\chi_\nu^2$ metric is estimated as follows:
\begin{equation} \label{eq_chi2}
\chi^2_{\nu} = \frac{1}{\nu}\sum_{i=1}^{N_{\rm data}} \frac{\left[ F_{i,\, {\rm data}}- A_p\cdot F_{i,\, {\rm model}}(\vec{p})\right]^2 }{ \sigma_{i,\, {\rm data}}^2}
,\end{equation}
where $N_{\rm data}$ is a number of pixels used to fit the data, $F_{i,\, {\rm data}}$ is the flux measured in pixel $i$ with an uncertainty $\sigma_{i,\, {\rm data}}$, and $F_{i,\, {\rm model}}(\vec{p})$ is the modeled flux of the same pixel. The degree of freedom of the fit is denoted by $\nu = N_{\rm data}-N_{\rm par}$, where $N_{\rm par}$ represents the number of model parameters $\vec{p}=(p_{1}, p_{2}, ..., p_{N_{rm par}})$.

To be less affected by single pixel noise and to speed up this procedure, we reduce the number of pixels of the original $Q_\varphi$ image (Fig. \ref{Qphi}(a)) by $8\times 8$ binning 
and select a rectangular image fitting area with a length of 63 
and width of 19 binned pixels (1.83$'' \times 0.55''$; Fig. \ref{model}(a)) centered on the star.
A round area with radius of 2.5 binned pixels covering the coronagraph 
and a few pixels around this region are excluded from the fitting procedure. 

The minimum $\chi^2_\nu =1.17$ is achieved for the model with parameters listed in Col.~4 of Table~\ref{t_results}. Considering all the models that fit the observations well and have a $\chi^2_\nu < 2,$ we 
derive the parameter distribution and fit a Gaussian for each model parameter (see Fig.~\ref{Hist}). We adopt the mean values of the parameter distribution 
as the best-fitting model presented in Col.~5 of Table~\ref{t_results}. The errors on the mean parameters are given by the  68~\% confidence interval derived for the respective distribution.

\subsubsection{MCMC modeling} \label{s_MCMC modeling}
We use the parameters of the best-fitting model from the parameter grid (Sect.~\ref{s_Grid}) as a starting point for the MCMC run. To perform this analysis, we employed the Python module $ emcee$ by \cite{Foreman-Mackey2013}, which implements affine invariant MCMC sampler suggested by \cite{Good2010}. The uniform priors are defined as specified in Col.~6 of Table~\ref{t_results}. Contrary to the modeling with a grid of parameters (Sect.~\ref{s_Grid}), the PA $\theta_{\rm disk}$ is allowed to vary. 

For the run of the standard ensemble sampler, 1000 walkers are initialized to perform a random walk with 2000 steps. The posterior distributions of the parameters shown in Fig.~\ref{f_mcmc} are obtained discarding the burn-in phase of 400 walker steps. The mean fraction of the accepted steps within the chain is 0.342 and the maximum autocorrelation time of parameter samples is 90 steps. This indicates a good convergence and stability of the chain by the end of the MCMC run \citep{Foreman-Mackey2013}. 

A set of the most probable parameters, which are considered to be the parameters of the best-fitting MCMC model, is listed in Col.~7 of Table~\ref{t_results}. These parameters are derived as the 50th percentile of the posterior distribution. Their lower and upper uncertainties are quoted based on the 16th and 84th percentiles of the samples, respectively. Figure~\ref{f_mcmc} demonstrates the degeneracy between some of the model parameters, for instance, between HG parameter $g_{\rm sca}$ and inclination of the disk $i$ or scale height $H_0$.

\subsubsection{Modeling results} \label{s_modeling results}
The MCMC sampling result for the PA of the disk is essentially the same as the PA we found by subtracting one disk side from the other (Sect.~\ref{s_PA}). Regarding other parameters, the best-fitting grid model (Table \ref{t_results} Col. 5) and the MCMC model (Table \ref{t_results} Col. 7) are consistent with each other. They indicate a disk radius $r_0$ between 10 and 12 au and a disk inclination of $\approx$103.5$^\circ$. The scale height of the vertical profile $H_0$ is 10 to 20 times smaller than the radius of the disk. According to the MCMC model, the outer density falloff is at least steeper than about $\alpha_{\rm out}= -7$. It is not clear whether the disk is cleared inside $r_0$ because the inner radial index $\alpha_{\rm in}$ is smaller than 3. Both models indicate a relatively high asymmetry parameter for scattering $g_{\rm sca} \approx 0.7$. This result could be affected by a higher surface brightness, which we observe inside $r=0.2''$. The flare index $ \beta$ is the only parameter that remains unconstrained in both modeling methods (grid and MCMC) showing a tendency to be 0.

Both models fit the $Q_\varphi$ image equally well according to the $\chi^2_{\nu}$ metric: the grid model has a $\chi^2_{\nu} = 1.20$ ($\nu = 1152$) and the MCMC model has a $\chi^2_{\nu} = 1.19$ ($\nu = 1151$). The residual images of both models show no noticeable differences. Therefore, in Fig.~\ref{model}, we show only the grid model in comparison to the $Q_\varphi$ image. 

Also, the radial profiles for the polarized flux of both models, which are calculated in the same bins as in the $Q_\varphi$ image (Fig.~\ref{model}(a)), are qualitatively very similar (cf. yellow solid and dotted lines in Fig.~\ref{f_SB_profile}a). 

\subsubsection{Modeling cancellation effect} \label{s_cancelation effect}
The synthetic image of the best-fitting model (Fig. \ref{model}(b)) 
is important for a correction of the polarized flux due to
the extended instrument PSF. One effect is the loss of
polarization signal because parts of it are outside the
flux integration aperture. The second and more important effect
for compact sources is the polarization cancellation
effect where the positive and negative $Q$ flux features in the Stokes 
$Q$ image, and similar for Stokes $U$, cancel each other substantially
if their separation is not much larger than the PSF. As described in
\cite{Schmid2006} and \cite{Avenhaus2014}, this can easily cause a reduction
of the measurable polarization flux by a factor of two or more,
depending on the source geometry and PSF structure.

We determine the total polarized flux for our best-fit model
before and after convolution with the instrument PSF and derive
for the cancellation effect a factor of 0.61 for our observations of the HD 172555 
debris disk. In addition
we find that only 82~\% of the polarized flux of the
convolved images fall into the rectangular disk-flux measuring
areas indicated by the white corners in Fig. \ref{Qphi}, while
4~\% of the flux is covered by the coronagraphic mask near the
star and 14~\% of the polarized flux are distributed in a very low surface brightness 
halo further out.

\begin{figure*}
\centering
    \includegraphics[width=17cm]{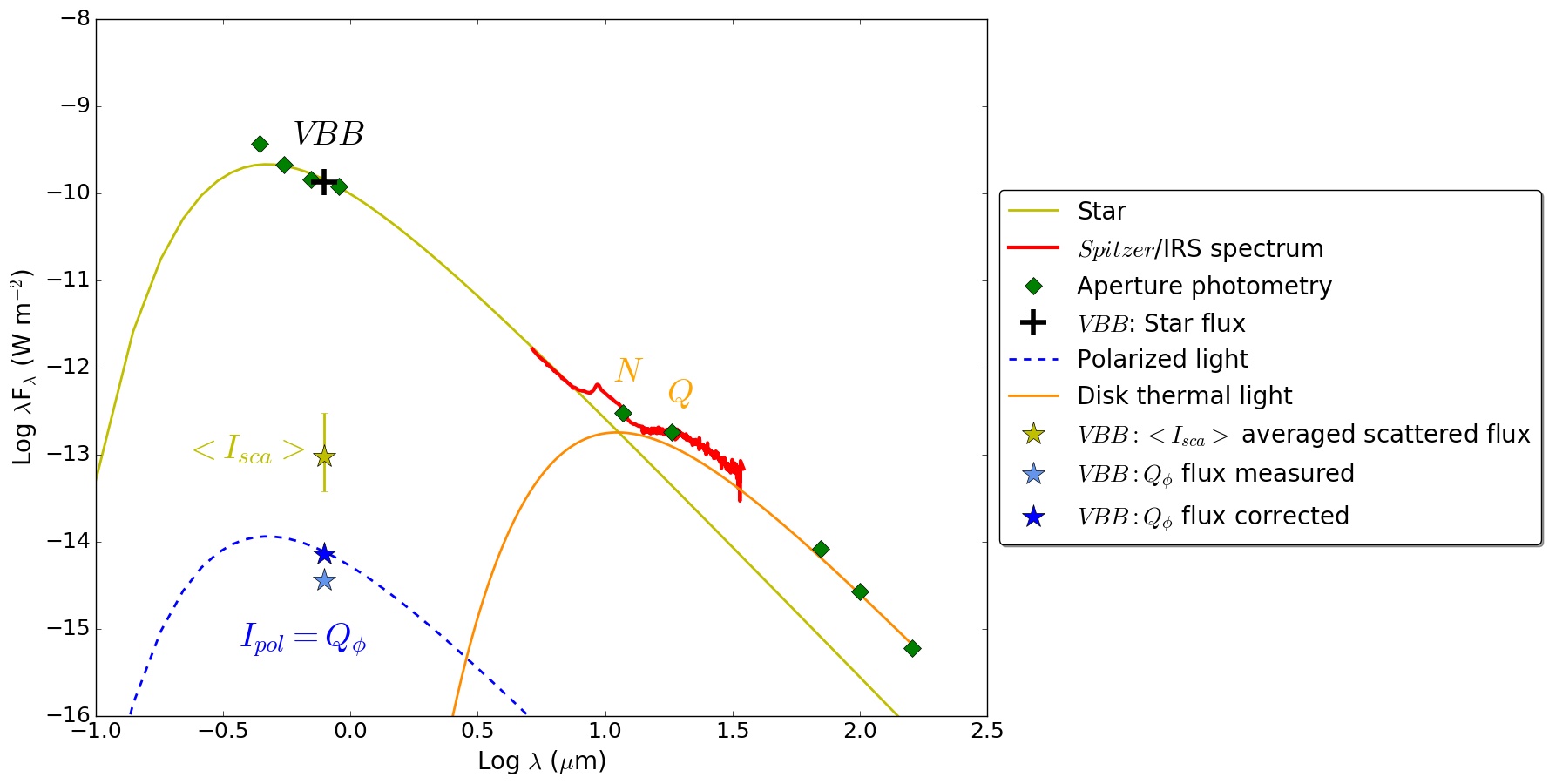} 
    \caption{Photometry of the star and disk together with the blackbody SED fits. \label{f_SED_LF}}
\end{figure*}

\subsection{Measured polarized flux contrast of the disk} 
\label{Contrast}
To compare the polarized flux from the disk with the stellar flux we
first derive the expected count rates of the star in the 
coronagraphic image, if there were no coronagraph hiding 
its PSF peak. This is obtained with the flux frames where the star 
is offset from the coronagraphic mask. We use two flux measurements 
from the nights when we obtained good results (177 and 249) and derive a mean 
count rate for the central star of $(4.36 \pm 0.08)\cdot10^5$ counts per second (ct/s) per ZIMPOL arm by summing up all counts registered 
within the aperture with a radius of $1.5''$ (413 pixels). Then we correct with a factor of 1.007 for the 
coronagraph attenuation at a distance of 0.35$''$ from the stellar 
PSF peak and a factor of $\sim$115 
to account for the mean transmission of the neutral density filter ND2.0 
in the VBB filter. We obtain an average count rate of 
$(5.00\pm 0.14)\cdot10^7$ ct/s per ZIMPOL arm for the expected
flux of the central star for observation in the fast polarimetry mode with the pupil mask but without correction for the attenuation by the focal plane mask or ND2.0 filter. 

For the disk, we derive the total polarized flux by summing up all the counts in two rectangular areas 
indicated with white brackets in Fig. \ref{Qphi}(a), from $|x|=0.08''$ to 0.77$''$ along the major axis and 
from $y= -0.13''$ to 0.16$''$ perpendicular to it. 
The innermost regions with radial separations $|x|<0.08''$
are largely hidden behind the coronagraphic mask, and therefore
they are not included in this estimate. The total polarized flux 
in the VBB filter after the modulation-demodulation efficiency calibration amounts to $780\pm150$ (ESE side, left) 
and $760 \pm150$ (WNW side, right) ct/s and per ZIMPOL 
arm. Therefore, the total net flux
obtained by integrating over abovementioned areas is $1540\pm300$ ct/s per ZIMPOL 
arm. \\

This measured flux must still be corrected for the polarimetric
cancellation effects to get a value for the intrinsic polarized 
flux. Because of the limited spatial resolution of 
our data, the polarization in the  
positive and negative $Q$ regions is reduced since signals
with opposite signs overlap and cancel each other. We quantified this effect and calculated
correction factors using our best-fit model (see description in the last paragraph of Sect. \ref{Modelling}).

Taking into account the PDI efficiency, which causes the reduction of disk polarized flux by 
a factor of $\sim$1.7 and correction factor 
for the aperture size (see Sect. \ref{Modelling}), the total
polarized flux of the disk is at least twice as large as the total net flux 
and is equal to $3100\pm600$ ct/s per ZIMPOL arm. 
This flux corresponds to the intrinsic modeled polarized flux compatible with our observation. With this value, the lower limit for the ratio of the disk total polarized flux to the 
stellar flux is $(F_{\rm pol})_{\rm disk}/F_{\rm \ast} \geqslant (6.2 \pm 0.6)\cdot 10^{-5}$ (cf. with the ratio $(F_{\rm pol})_{\rm disk}/F_{\rm \ast} = 3.1\cdot 10^{-5}$ obtained for the net flux $(F_{\rm pol})_{\rm disk} = 1540$ ct/s).

\subsection{Polarimetric flux} 
\label{s_Star_flux}
To check our photometry, we compare the measured stellar flux with the stellar magnitudes from the literature.
The stellar count rate of $(4.36\pm 0.08)\cdot10^5$ ct/s per ZIMPOL arm before 
correction for the transmission of the neutral density filter
can be converted to the photometric magnitude $m$(VBB) \citep{Schmid2017} as follows:
\[m(\mathrm{VBB}) = -2.5 \log (\mathrm{ct/s})-\mathrm{am}\cdot k_1(\mathrm{VBB})-m_{\mathrm{mode}}+ \mathit{z}p_{\mathrm{ima}}(\mathrm{VBB}) ,\]
where am~$=1.3$ is the airmass, $ k_1(\mathrm{VBB})=0.086^m$ is 
the filter coefficient for the atmospheric extinction, 
$\mathit{z}p_{\mathrm{ima}}(\mathrm{VBB})=24.61^m$ is the photometric 
zero point for the VBB filter, and $m_{\mathrm{mode}}= 5.64^m$ is an offset 
to the zero point, which accounts for the instrument configuration, pupil stop (STOP1\_2), neutral density filter ND2.0, and the fast polarimetry detector mode.
For HD 172555 we derive a magnitude 
$m(\mathrm{VBB})$ = 4.68$^m \pm 0.03^m$ (indicated with a black asterisk in Fig.~\ref{f_SED_LF}). 
This value is in good agreement with the Johnson photometric 
magnitude in R band $m(\mathrm{R})$ = $4.887^m \pm 0.023^m $ 
and I band $m(\mathrm{I}) = 4.581^m \pm 0.025^m $ \citep{Johnson1966}.

For the estimated above ratio of the disk total polarized flux to 
stellar flux $(F_{\rm pol})_{\rm disk}/F_{\rm \ast} \geqslant (6.2 \pm 0.6)\cdot 10^{-5}$, disk magnitude in the VBB filter is $mp_{\mathrm{disk}}(\mathrm{VBB})$ = 15.20$^m \pm$ 0.37$^m$ (indicated with a blue asterisk in Fig.~\ref{f_SED_LF}). 

At radial separation of $r \approx 0.3''$ indicated with arrows "A" and "B" in Figs.~\ref{Qphi}(b) and \ref{f_SB_profile}, the peak surface brightness is $\sim$15~ct/s per binned pixel (0.029$'' \times 0.029''$) and corresponds to the ${\rm SB} \mathrm{_{peak}(VBB)} = 13.3^m \pm 0.3^m$ arcsec$^{-2}$ or surface brightness contrast for the polarized flux of ${\rm SB} \mathrm{_{peak}(VBB)} - m \mathrm{_{star}(VBB)} = 8.62$ mag arcsec$^{-2}$. 
 
\section{Discussion} \label{Discussion}
\subsection{Comparison with thermal light detection and interferometry} \label{s_Smith}

The disk geometry is consistent with the $18~\mu$m image presented in \cite{Smith2012}. 
They found two lobes of extended emission along PA $= 110^\circ$
with a separation of about $0.4''$ from the star and a flux of 
$105$~mJy. On top of this, they measured a
roughly seven times stronger (732~mJy) unresolved dust component and
a stellar flux of 202~mJy. In the N-band image the dust was 
not resolved by \citet{Smith2012}, but their N-band MIDI 
interferometry resolved the dust fully, putting it at 
separations between $>0.035''$ and $<0.27''$ ($1-8$~au). Therefore,
we also expect that a lot of polarized light from the dust
scattering contributes to the signal inside the inner working angle 
of the SPHERE/ZIMPOL observations at $0.12''$, which cannot be measured. 

The disk modeling of the infrared flux by \citet{Smith2012} yields 
a disk with radius $r=0.27''$ and width $dr= 1.2r,$ which is 
equivalent to the ring with constant surface brightness between $0.1''$ and 
$0.4''$, inclined at 75$^\circ$ to the line of sight 
and at PA$= 120^\circ$ for the disk major axis. We confirm these
parameters with our observation and can provide more stringent parameters for the inclination
and the disk orientation because of
the higher spatial resolution of our data. Also the disk extension
agrees, but it is not clear whether the scattered light emission
and thermal emission should show the same radial flux distributions.

In Fig.~\ref{f_SED_LF}, we compare the flux distribution $\lambda F_{\lambda}$ from the stellar photosphere with the thermal emission of the disk and the polarized flux distribution measured in this work. 
Photometric data points of HD 172555 are listed in Table~\ref{t_SED} and plotted as green diamonds in Fig.~\ref{f_SED_LF}, which also includes a $Spitzer$/IRS spectrum\footnote{downloaded from http://www.stsci.edu/~cchen/irsdebris.html}. The stellar spectral flux density is approximated by a Planck function for the star temperature of 7800 K \citep{Riviere-Marichalar2012}.
A black asterisk (labeled VBB) denotes the stellar magnitude measured in the VBB (this work). The near-IR aperture photometry of HD 172555 in the N and Q bands (green diamonds at 11.7 and 18.3 $\mu$m), performed by \citet{Smith2012} using TReCS imaging data, is indicated by capital letters `$N$' and `$Q$'. The thermal flux from the disk is shown as a blackbody emission with the temperature of 329 K found by \cite{Riviere-Marichalar2012} (orange solid line).  
The net polarized flux measured in the $Q_\varphi$ image (Sect.~\ref{Contrast}) is indicated by a light blue asterisk (labeled with '$I_{pol}=Q_\varphi$'), and the polarized flux corrected for the polarimetric cancellation effects and aperture size is indicated by a blue asterisk in this figure. The blue dotted line shows an approximate curve for the SED of the polarized light obtained by scaling down the SED of the star. 
Using our best-fitting grid model we roughly estimated the scattered flux from the disk (averaged over full solid angle) in the VBB filter. This conversion neglects the contribution from the diffracted light and assumes that the maximum polarization fraction of the phase function for the polarized flux is $p_{\rm max}= 0.3$. This value is denoted with a yellow asterisk and labeled '<$I_{sca}$>'. We used this magnitude as a proxy for the scattered flux of the disk to compare it with the maximum thermal flux of the disk at $\lambda = 10$ $\mu$m and thus to estimate a kind of the disk albedo. We obtain <$I_{sca}$>$\,(p_{\rm max}= 0.3)$/$\lambda F_{\lambda}\,$($\lambda = 10$ $\mu$m) = 0.54, which should be considered as a very rough estimate for this ratio. The lower and upper limits on the scattered flux $I_{sca}$, shown in Fig.~\ref{f_SED_LF}, are obtained assuming maximum polarization fraction $p_{\rm max}= 0.1$ (for the upper limit) and $p_{\rm max}= 0.7$ (for the lower limit).  

\begin{table}  
     \caption[]{Photometry of HD 172555. \label{t_SED}}
             \centering
         \begin{tabular}{lcccc}
            \hline
            \hline
            \noalign{\smallskip}
       Filter & Wavelength & Flux density & Error & Ref.\\  
        & ($\mu$m) & (Jy) & (Jy) &  \\   
           \hline
            \noalign{\smallskip}
       Johnson $B$ & 0.44  & 53.87 & 0.94 & 1 \\[5pt]
           Johnson $V$ & 0.55  & 38.45  & 0.53 & 1 \\[5pt]
           Johnson $R$ & 0.7  & 33.40  & 0.71 & 1\\[5pt]
       Johnson $I$ & 0.9  & 35.74  & 0.82 & 1 \\[5pt]
       Si-5 ($N$) & 11.7 & 1.120  & 0.067 & 2 \\[5pt]
       Qa ($Q$) & 18.3 & 1.039  & 0.085 & 2 \\[5pt]
       PACS70 & 70 & 0.191 & 0.005 & 3 \\[5pt]
       PACS100 & 100 & 0.089 & 0.003 & 3 \\[5pt]
       PACS160 & 160 & 0.036 & 0.02 & 3 \\[5pt]
           \noalign{\smallskip}
           \hline
            \hline
            \noalign{\smallskip}
         \end{tabular}
\begin{flushleft}
{\bf References.} (1) \citet{Johnson1966}; (2) \cite{Smith2012}; (3) \cite{Riviere-Marichalar2012}. 
\end{flushleft}
\end{table}

\begin{figure*}
\centering
    \includegraphics[width=16cm]{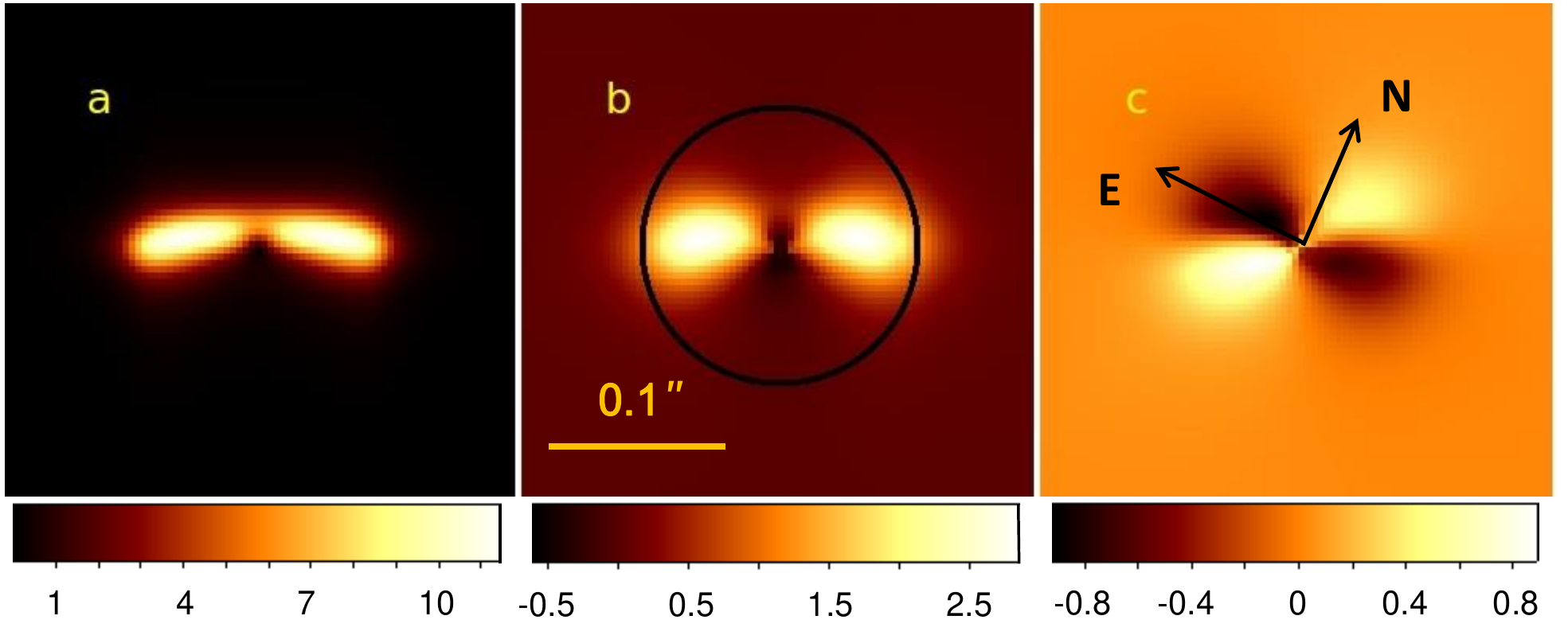} 
    \caption{Model of the small debris disk that could be hidden behind the coronagraphic mask. The model parameters correspond to those of the mean model (Table \ref{t_results}) except the radius of the belt and scale height of the dust vertical distribution, which are reduced by a factor of 5. ({\bf a}) $Q_\varphi$ image of the debris disk model showing the polarized intensity before being convolved with the instrumental PSF. ({\bf b}) $Q_\varphi$ image of the debris disk model showing the polarized intensity after convolution with the instrumental PSF. The black circle denotes the edge of the coronagraphic mask. ({\bf c}) $U_\varphi$ image of the model demonstrating nonzero $U_\varphi$ signal appearing after convolution of the Stokes $Q$ and $U$ parameters with the instrumental PSF. The color bars show counts per pixel. \label{f_hidden}}
\end{figure*}

\subsection{Hidden source of thermal emission }
According to the mid-IR observations of \cite{Smith2012},  
most of the dust is located inside 8~au, down to 1~au.  
Therefore, we investigate with model calculations whether 
a strong but compact source of scattered light could be present, 
which is not detectable in our data because of the limited
resolution and inner working angle limit of the used coronagraph. 

The occulting spot in our coronagraphic data has a radius 
$r\sim0.08''$ (2.27 au). We model a small disk with a radius 
of 2 au (0.07$''$) with exactly the same geometric morphology
as the best-fit grid model derived in Sect~\ref{Modelling},  but 
just smaller by a factor of 5. This small disk is shown als $Q_\varphi$ image
in Fig.~\ref{f_hidden}(a), together with the convolved $Q_\varphi$ image 
with the size of the coronagraphic mask indicated in Fig.~\ref{f_hidden}(b). 
First, we notice a large difference between $Q_\varphi$
intrinsic polarized flux and the convolved polarized flux of a factor
of 2.5 because of the very strong polarimetric cancellation 
for such a compact disk. Then, there is another factor 2.67 
between the total convolved polarized flux from the disk and the 
convolved polarized flux that falls into the rectangular 
measuring areas of our observations shown in Fig.~\ref{Qphi}(a)
because a major region of the polarization signal is
hidden by the coronagraphic mask. Thus, only about 15~\% of the
polarized flux produced by an inner compact disk would contribute to
our measurement. We measure a net flux of 1540 ct/s in the
measuring area of our observation and estimate that it is 
not possible to recognize an inner disk in our data, 
which contributes less 
than 150 ct/s to our measurement. Thus, an unseen compact disk with
the geometry as described above and 
an intrinsic polarized flux of 6.675 x 150 ct/s could be present 
without being in conflict with our observations and even more scattered 
light could be hidden for a more compact inner disk.

\subsection{Comparison between polarized flux and thermal emission} \label{s_Lambda parameter}

To characterize the scattering albedo of the dust, we compute the $\Lambda$ parameter describing the ratio of the fractional polarized light flux to fractional infrared luminosity excess of the disk \citep{Engler2017}, i.e.,
\begin{displaymath}
\Lambda = {(F_{\rm pol})_{\rm disk}/F_{\rm \ast} \over L_{\mathrm{IR}}/L_{\ast}},
\end{displaymath}  
where the ratio of total polarized flux of the disk to the stellar flux for HD 172555 is $(F_{\rm pol})_{\rm disk}/F_{\rm \ast} \geqslant (6.2 \pm 0.6)\cdot 10^{-5}$ (Sect. \ref{Contrast}). Adopting the ratio of the disk infrared luminosity to stellar luminosity $L_{\rm {IR}}/L_{\ast}$ equal to $7.2\cdot 10^{-4}$ \citep{Mittal2015}, we obtain a lower limit for $\Lambda \geqslant 0.086$. This value is slightly smaller than the $\Lambda$ parameters we have estimated for the F star HIP 79977 and the M star AU Mic ($\Lambda_{\rm HIP\, 79977}=0.11$ and $\Lambda_{\rm AU\, Mic}=0.55$) in \citet{Engler2017}.

\section{Summary} \label{s_Summary}
In this paper, we presented images of polarized scattered light 
from the debris disk around HD 172555 obtained in the VBB filter 
using differential polarimetry. We found that the 
observed polarized intensity is consistent with an axisymmetric dust distribution, 
which can be interpreted as a parent belt of planetesimals 
or disk with a radius in the range between 0.3$''$ (8.5 au) and 0.4$''$ (11.3 au). We analysed the disk structure
and obtained the following results:
\begin{itemize}
\item The PA for the disk major axis is $\theta_{\rm disk} = 112.3^{\circ} \pm 1.5^{\circ}$ on the sky.
\item The disk emission is slightly shifted in NNE direction from the star indicating
that the front side or forward-scattering part of the disk is on the NNE side. 
The observed dust distribution can be described with the 
HG asymmetry parameter $g\approx0.7$ and disk inclination $\approx103.5^{\circ}$.
\item Data analysis and modeling results do not suggest the clearing of 
the inner disk regions (inside $r=0.3''$).

\item The total disk magnitude in polarized flux in the VBB filter is 
$mp_{\mathrm{disk}}(\mathrm{VBB})$ = 15.20$^m \pm$ 0.37$^m$ and the stellar flux is 
$m(\mathrm{VBB})$ = 4.68$^m \pm 0.03^m$. 
This gives a disk to star contrast $(F_{\rm pol})_{\rm disk}/F_{\rm \ast}$ of $(6.2 \pm 0.6)\cdot 10^{-5}$. The measured peak surface brightness of the polarized light is ${\rm SB} \mathrm{_{peak}(VBB)} = 13.3^m \pm 0.3^m$  arcsec$^{-2}$. This corresponds to a surface brightness contrast of 
${\rm SB} \mathrm{_{peak}(VBB)} - m \mathrm{_{star}(VBB)} = 8.62$ mag arcsec$^{-2}$. 
\item When compared with the fractional infrared luminosity of the disk using 
the $\Lambda$ parameter, the fractional polarized light flux in the VBB filter 
makes up $\sim 9\%$. 
\end{itemize}
Our data demonstrate high sensitivity and ability of ZIMPOL to resolve the polarized light from
the hot debris around a nearby A star as close as $0.1''$ or $\sim$3 au. It would be worth reobserving HD172555 
in the I band without the coronagraph. Additional 
observations of this target under excellent observing conditions, i.e., seeing $\leqslant 0.7$,
airmass $\leqslant 1.4$, coherence time $ \geqslant 3.5$ ms, and wind speed $>3$ m/s, to avoid 
the low wind effect (see SPHERE Manual) and with different offsets of the sky field on the detector 
would allow us to better distinguish between the instrumental features in the PSF
and the real astrophysical signal. In the VBB filter the speckle ring extends from
$0.30''$ to $0.45''$ and overlaps the region of interest. In the I band, 
the speckle ring is further outside of $\sim0.40''$ and the achievable contrast is higher under 
optimal observing conditions even if the amount of photons received in the I band is a factor of 0.6 lower 
than in the VBB. Moreover, noncoronagraphic data taken with I band are more suitable for the
measurement of the beam shift between the two orthogonal polarization states and, hence, the correction of the beamshift effect (Schmid et al. 2018, submitted) is much easier. Such observation may also allow us to probe the innermost dust in the HD 172555 system.

\begin{acknowledgements}
We would like to thank the referee for many thoughtful comments that helped to improve this paper. This work is supported by the Swiss National Science Foundation through grant number 200020 - 162630.
\end{acknowledgements}

\bibliographystyle{aa} 

\bibliography{reference.bib} 

\newpage 
\appendix

\section{Modeling the $Q_\varphi$ and $U_\varphi$ images} \label{s_modeling_app}
 \begin{figure*}
\centering
    \includegraphics[width=17.5cm]{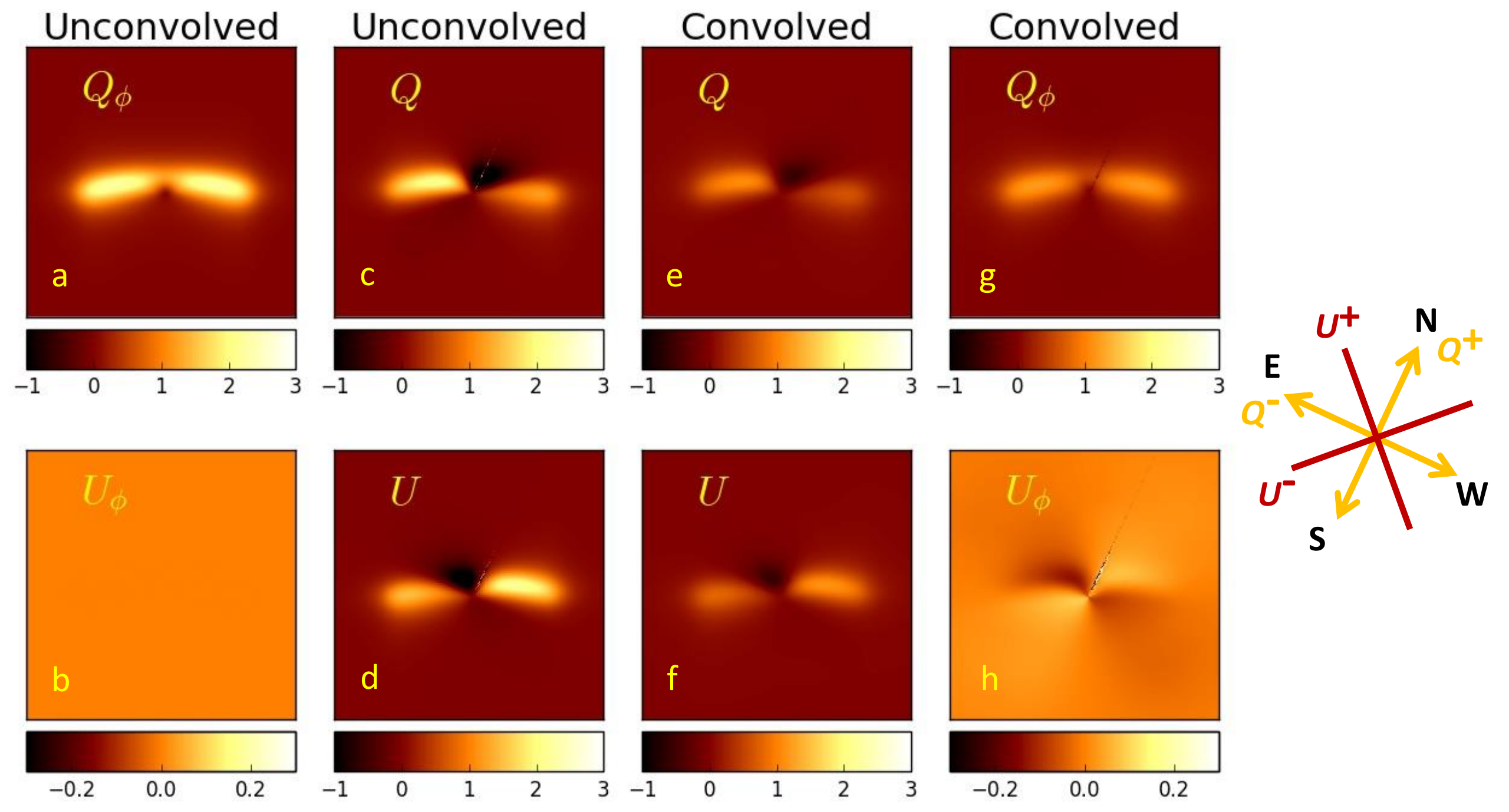} 
    \caption{Model images of the polarized flux from HD 172555 debris disk. Images were generated using the parameters of mean model (Col. 5 in Table \ref{t_results}) and illustrate the individual steps of the modeling procedure (from left to the right) to obtain the final synthetic $Q_\varphi$ and $U_\varphi$ images (({\bf g}) and ({\bf h})). For a detailed description of the presented images, see Appendix \ref{s_modeling_app}. The diagram on the right side of the figure explains orientation of the axes used to measure the Stokes $Q$ and $U$ parameters as implemented in ZIMPOL instrument. The diagram shows also the orientation of north and east in each image. The north is rotated by 22$^\circ$ clockwise from the vertical line to have disk axis in horzontal position. Nonalignment of the disk axes with the $Q$ and $U$ measuring axes causes asymmetric flux distribution in $Q$ and $U$ images (({\bf c}), ({\bf d}), ({\bf e}), and  ({\bf f})). \label{f_app_conv_effect}}
\end{figure*} 

Figure \ref{f_app_conv_effect} simulates the polarized flux measurement with the ZIMPOL instrument. In the first step, the distribution of the dust number density in the disk is generated using the parameters of the mean model. Then, synthetic images of debris disk in polarized light are calculated assuming only the single scattering of photons.
An image of the polarized imtensity, or $Q_\varphi$ image (Fig. \ref{f_app_conv_effect} (a)), is obtained considering angular dependence of polarization fraction on scattering angle $\theta$ such as that for Rayleigh scattering. Single scattering of photons produces linearly polarized light oriented perpendicular to the scattering plane, which corresponds to the azimuthal orientation of electric field in the 2D $Q_\varphi$ image. The $U_\varphi$ image (Fig. \ref{f_app_conv_effect} (b)) displays no signal because the Stokes $U_\varphi$ parameter should show the polarization component in direction at 45$^\circ$ to the azimuthal direction. In this direction we do not expect to measure any signal if the assumption of single scattering is valid. 

In the following step, the Stokes $Q$ and $U$ images are calculated using the coordinate transformation
\begin{equation}
Q = - Q_\varphi \cos 2\varphi
\end{equation}
\begin{equation}
U = - Q_\varphi \sin 2\varphi
,\end{equation}
where $\varphi$ is the polar angle of each image point and is measured from NoE. This polar coordinate system corresponds to the measurement axes of the instrument shown on the right side of Fig. \ref{f_app_conv_effect}. In ZIMPOL, the Stokes $Q^+$ parameter is measured along north-south axis and the Stokes $U^+$ parameter is measured along northeast-southwest axis.

The $Q$ and $U$ images (Figs. \ref{f_app_conv_effect} (c) and (d)) are employed to create the images of individual polarization states $I_0$, $I_{45}$, $I_{90}$, and $I_{135}$. These images are firstly convolved with the PSF measured from the HD 172555 data and then are utilized to calculate new images of the $Q$ and $U$ parameters again but now in convolved state (Figs. \ref{f_app_conv_effect} (e) and (f)). These two images imitate the output of the ZIMPOL data reduction pipeline. Usually they are used to compute the final $Q_\varphi$ and $U_\varphi$ images of the disk applying reverse coordinate transformation as follows:\begin{equation}
Q_\varphi = - (Q\cos 2\varphi+U\sin 2\varphi),
\end{equation}
\begin{equation}
U_\varphi = -Q\sin 2\varphi+U\cos 2\varphi
.\end{equation}

Examination of the final $Q_\varphi$ and $U_\varphi$ images (Figs. \ref{f_app_conv_effect} (g) and (h)) leads to two important conclusions. Comparison of the unconvolved and convolved $Q_\varphi$ images (Figs. \ref{f_app_conv_effect} (a) and (g)) demonstrates the loss of a fraction of the polarized flux due to the convolution of the intensity of two orthogonal polarization states with the instrumental PSF (see also Sect.~\ref{Modelling}). As a result of blurring effect on the positive and negative signals in the Stokes $Q$ and $U$ due to the convolution with PSF, a nonzero signal is present in the final $U_\varphi$ image (Fig. \ref{f_app_conv_effect} (h)). For the mean model shown in Fig.~\ref{f_app_conv_effect}, the generated $U_\varphi$ flux is of one order smaller than the $Q_\varphi$ flux that we measure in the $Q_\varphi$ image (Fig. \ref{f_app_conv_effect} (g)). In general, this artificially created $U_\varphi$ flux and the magnitude of the flux loss in the $Q_\varphi$ image depend on the geometry and extent of the polarized flux source and on the PSF structure. For the small compact sources, the generated $U_\varphi$ flux and the PDI efficiency loss are the largest because the resolution of instrument is limited.
Therefore, this effect should be estimated by means of modeling for each individual object and detector. To demonstrate this point, we calculate two additional models with the same parameters and the same aspect ratio of radius of the planetesimal belt to scale height as in the mean model but with different radii of the belts: one model with radius $r=2$ au ($0.07''$) shown in Fig. \ref{f_hidden} and another model with the radius $r=30$ au ($1.05''$). Our results show that the total polarized flux of the smaller debris disk is reduced by a factor of 0.4, whereas the extended disk suffers much less from the convolution effect and reducing factor is 0.73 for this disk.

\newpage 
\section{HD 172555 polarimetric data per observing run}
\begin{SCfigure*}
\caption{Reduced imaging data of HD 172555 from all six observing runs arranged from top to the bottom with the Stokes $I$ (left column), $Q$ (middle column) and $U$ (right column) data. 
Each run was carried out with fixed sky orientation on the CCD detector with PA offset indicated in parenthesis. The highest quality observations are in rows two and four. To get north up and east to the left, each image should be rotated by the PA offset counterclockwise and the True North offset of the ZIMPOL should be taken into account. The yellow lines in the $Q$ and $U$ images show the position of disk major axis. 
\label{IQU}  }
 \includegraphics[width=12cm]{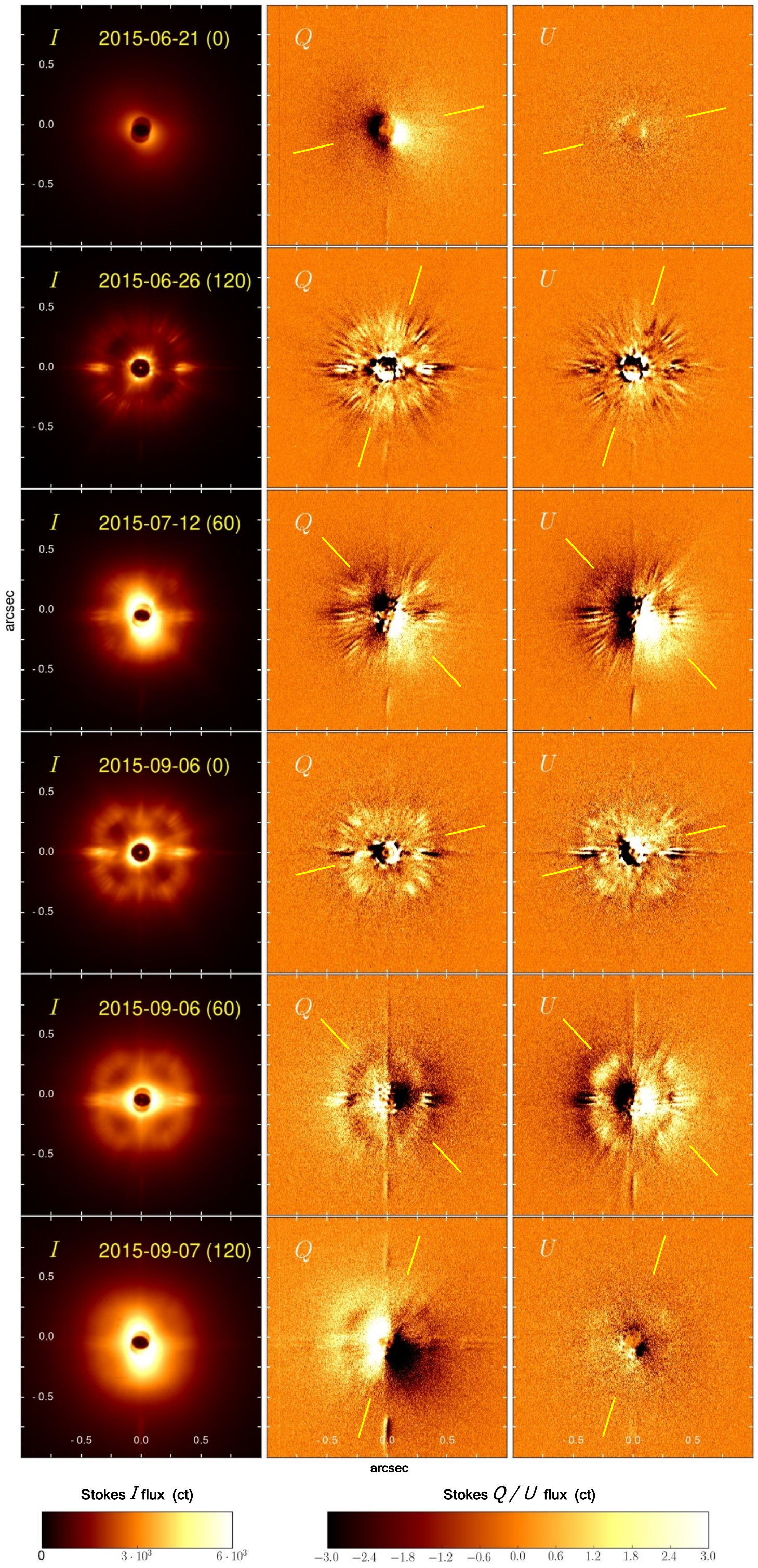} 
 \end{SCfigure*}

Figure~\ref{IQU} shows the reduced imaging data of HD 172555 from all six observing runs. The FITS files of the total intensity (Stokes $I$) and Stokes parameters $Q$ and $U$ of the data presented in rows two and four in this figure are available in electronic form at the CDS via anonymous ftp to cdsarc.u-strasbg.fr (130.79.128.5) or via http://cdsweb.u-strasbg.fr/cgi-bin/qcat?J/A+A/. 

\section{Probability distributions for fitted parameters}
\begin{figure*}
   \centering
  \includegraphics[width=18cm]{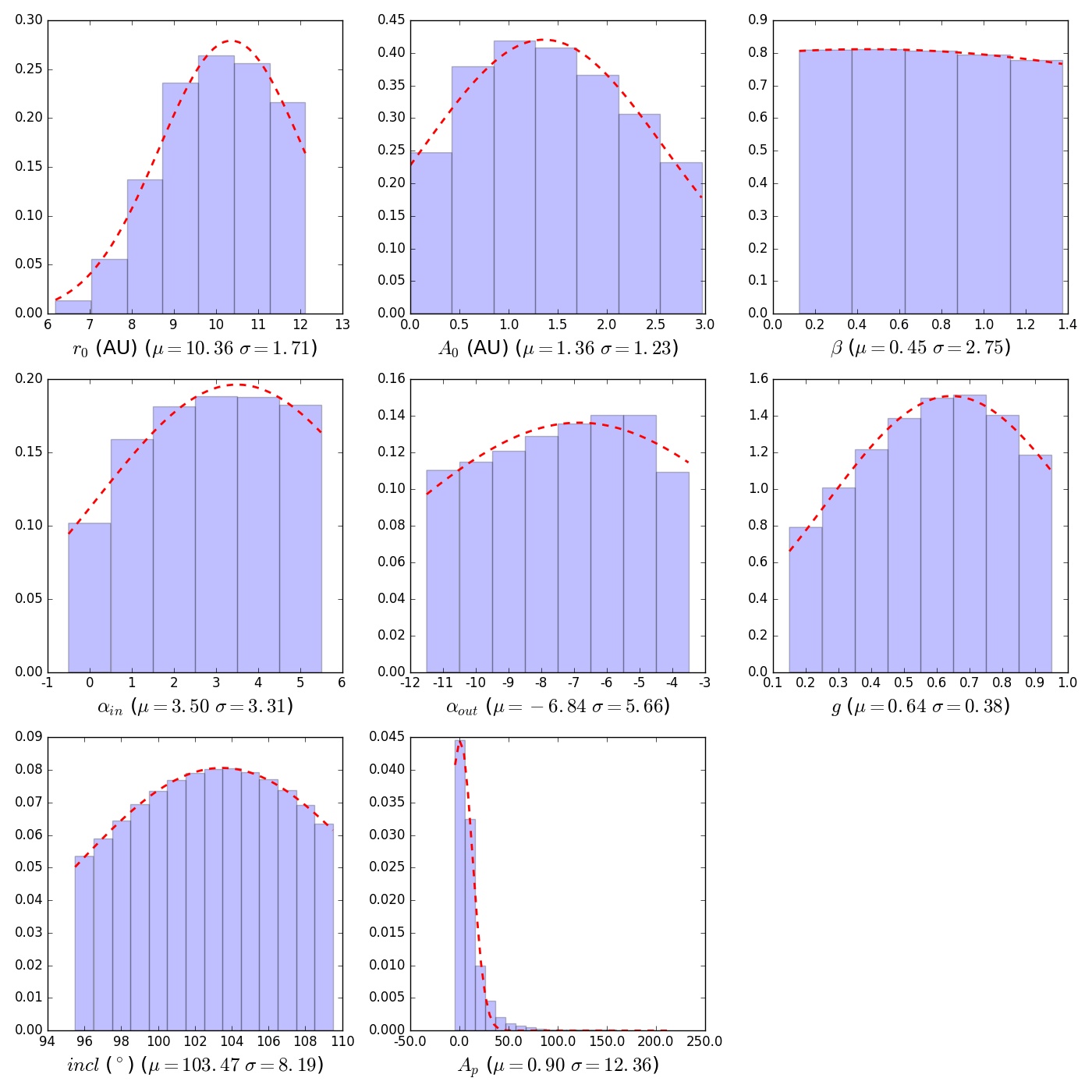} 
    \caption{One-dimensional marginalized distributions of model parameters drawn from the sample of well-fitting models ($\chi^2_\nu < 2$). The red dotted lines denote Gaussian fits to the parameter distributions. Their mean values $\mu$ and standard deviation $\sigma$ are given in brackets. }
               \label{Hist}
   \end{figure*}

\begin{figure*}
   \centering

      \includegraphics[width=18cm]{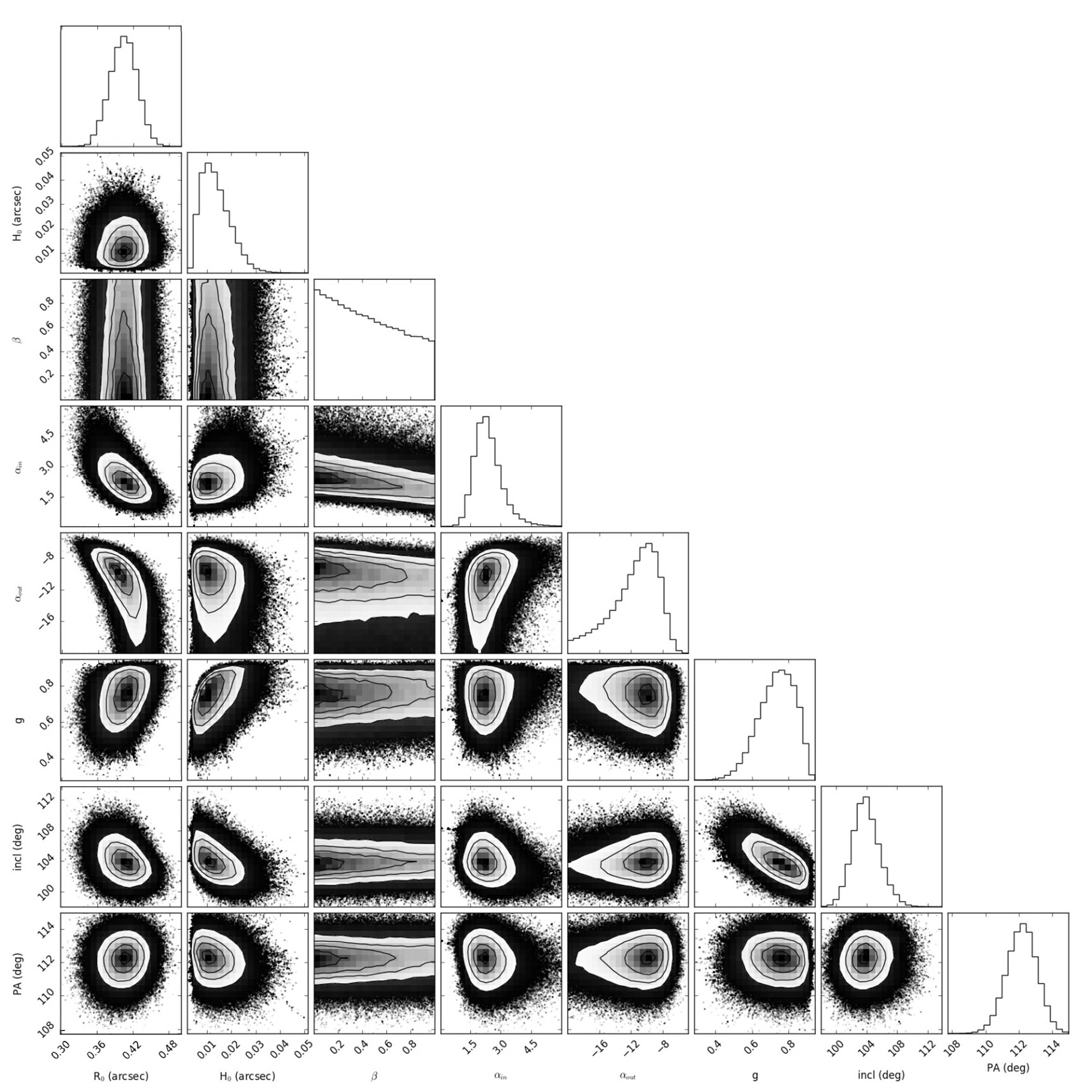}   
     \caption{One- and two-dimensional posterior probability distributions of the fitted parameters.}
               \label{f_mcmc}
\end{figure*}

See Figures~\ref{Hist} and \ref{f_mcmc}.

\end{document}